\documentclass[a4paper,10pt,dvips]{article}
\usepackage{amsmath}
\usepackage{amssymb}
\input{epsf.sty}
\newcommand{\miosinadue}{Myosin~\mbox{II}}
\newcommand{\cosbol}{$k_{\mbox{\tiny B}}$}
\newcommand{\eneter}{$k_{\mbox{\tiny B}}T$}
\newcommand{\numdim}[2]{$#1$~#2}

\newcommand{\cvd}{\par\raggedleft{\rule{2mm}{2mm}}}
\newenvironment{figMacPc}[4]
{
\begin{figure}[htb]
   \hfill
         \begin{minipage}[htb]{#2}
            \epsfxsize=#2\centerline{\epsfbox{#1}} 
            \caption{#3}                            
            \label{#4}
         \end{minipage}                          
   \hfill
}
{
   \end{figure}
}
\newenvironment{tabMacPc}[4]
{
   \begin{table}[htb]
   \hfill      
         \begin{minipage}[htb]{#2}  
            \caption{#3}                            
            \begin{center}{#1}\end{center}
            \label{#4}                         
         \end{minipage}                          
   \hfill
}
{
   \end{table}
}
\newenvironment{acknowledgment}
{
  \section*{Acknowledgment}}{\par\addvspace{12pt}
}

\title
{{Lever-arm and Washboard-potential theories\\
jointly account for \miosinadue\ dynamics}\thanks{Work partially supported by MIUR (PRIN 2003) and by Campania Region.}}
\author
{
\\[0.8cm]
A. Buonocore, L. Caputo, E. Pirozzi and L. M. Ricciardi\\[0.8cm] 
Dipartimento di Matematica e Applicazioni\\
Universit\`{a} di Napoli Federico II\\
{\{aniello.buonocore, enrica.pirozzi, luigi.ricciardi\}@unina.it}\\
luigia.caputo@dma.unina.it
}
\date{}
\begin{document}

\maketitle
\begin{abstract}
We address the controversial hot question concerning the validity of the loose-coupling versus the lever-arm models in the actomyosin dynamics by re-interpreting and extending the washboard potential model proposed by some of us in a previous paper. In the new theory, a loose-coupling mechanism co-exists with the deterministic lever-arm model. The synergetic action of a random component, originating from the harnessed thermal energy, and of the power-stroke generated by the lever-arm classical mechanism is seen to yield an excellent fit of the set of data obtained in T. Yanagida's laboratory on the sliding of Myosin II heads on actin filaments under various load conditions. Our theoretical arguments are complemented by accurate numerical simulations, and the robustness of theory is tested via different combination of parameters and potential profiles.
\end{abstract}
\vspace*{1cm}
\section{Introduction}
In 1985 Yanagida \emph{et al.}~\cite{yan85} described a sophisticated experimental set-up to measure the distance traveled by an actin filament interacting with a \miosinadue\ head\footnote{Hereafter we shall use the word \lq\lq myosin\rq\rq\ as an abbreviation of \lq\lq\miosinadue\rq\rq.} during a complete ATP cycle. Under low load conditions on the actin filament, they could observe traveled distances of up to \numdim{60}{nm}. This was in strong disagreement with other experimental evidences (see~\cite{fin94},~\cite{meh97} and references therein), indicating instead distances ranging between 4 and 10 nanometers, such smaller range of traveled distances is coherent with the Lever-arm theory (see~\cite{spu94} and~\cite{coo97}) that is still widely believed to account for the generation of the force responsible for the actin filament sliding. The level-arm theory can be essentially schematized as follows (more details can be found in~\cite{myohp}, where a graphical animation is also present):
\begin{itemize}
\item[--]{The binding of an ATP molecule on the catalytic site of myosin head produces the detachment of the head from the actin filament.}
\item[--]{The ATP hydrolysis, that follows within about the next \numdim{10}{ms}, changes ATP to ADP.Pi with simultaneous release of energy in the amount of about \numdim{20}{\eneter}. ATP hydrolysis also generates a sensible rotation of the neck of the myosin head.}
\item[--]{The transition M.\,ATP$\longrightarrow$ M.\,ADP.\,Pi is accompanied by an increase of affinity of this complex for actin, which enhances the chance of myosin to get binded on an actin site. Since such binding is reversible, the myosin head can visit more than a single actin site.}
\item[--]{While myosin head is binded at an actin site, its neck may suddenly switch back to its original position (this is the so called \lq\lq power stroke\rq\rq) with the successive release of the phosphoric radical Pi and the occurrence of conformational changes around the myosin coiled coil\footnote{The nature of such conformational changes has not been yet fully understood. With reference to kinesin, a possible mechanism is described in~\cite{kik01} where its validity also for the myosin family is conjectured.}. This process is reversible i.e. phosphoric radical Pi can re-establish the complex M.\,ADP.\,Pi. In the absence of such recombination, the above-mentioned conformational changes are very likely to generate the sliding of the actin filament with the release of ADP molecule after about \numdim{2}{ms}.}
\item[--]{The myosin head stays then binded on the actin site till the arrival of a new ATP molecule, which starts the whole process afresh.}
\end{itemize}
\par
Summing up, as far as reference is made exclusively to the mechanical phenomenon of the generation of the force responsible for the movement, the Lever-arm theory is strictly deterministic: Each ATP cycle generates one single power-stroke that causes a sliding of constant length and preassigned direction of the actin filament. Within such framework, a tight coupling between ATP cycles and protein movements is envisaged.
\par
A dramatically different conclusion appears to be required by the data of Yanagida and co-workers as indicated in the sequel. Indeed, aiming at a continuous efforts towards increasingly accurate measurements of the traveled distances during an ATP cycle, they were able to achieve impressive improvements of the technological set-up, including the design and construction of \lq\lq home made\rq\rq\ highly sophisticated devices. As described in~\cite{kit99}, by exploiting such a technology, they proceeded as follows: A myosin head was attached to the tip of a glass microneedle and placed near an actin filament that had been previously immobilized on a microscopy slide by means of optical tweezers. The deflections of the needle with respect to its resting position were then measured and recorded. From the traces thus obtained the following three features emerged that are in evident disagreement with the tight coupling theory:
\begin{enumerate}
\item{The overall displacement traveled by the myosin head is not constant; it can be as large as \numdim{30}{nm}.}\label{Oss1}
\item{This distance is the sum of a random number of single \lq\lq steps\rq\rq, the amplitude of each of which equals the distance (\numdim{\simeq~5.3}{nm}) between two successive actin monomers. During the time elapsing between two successive steps, the myosin head randomly jitters around an equilibrium position.}\label{Oss2}
\item{Steps mainly occurs in a fixed \lq\lq forward\rq\rq\ direction, although some of them occasionally take place in the opposite \lq\lq backward\rq\rq\ direction. Hereafter, forward steps will be taken as positive and backward steps as negative. The overall displacement is thus the algebraic sum of the number of performed forward and backward steps.}\label{Oss3}
\end{enumerate}
Such evidence contrasts the one-to-one relation hypothesized between ATP hydrolysis and the occurrence of the mechanical event consisting of the power stroke in the myosin head. Furthermore the observation of the existence of random elements leads one to conjecture that a significant role could be played in such context by the thermal agitation of the environmental molecules of the watery solution in which the involved proteins are embedded. This is the motivation for the assumption of the existence of a loose coupling between ATP cycle and actomyosin dynamics (See, for instance, \cite{oos00}).
\par
While referring to~\cite{cyr00} for a lucid outline of the origin of the controversy existing among the supporters of the loose coupling mechanism and the community of those faithful to the Lever-arm theory, and therefore also the tight coupling vision, in the present paper we purpose to resolve such a controversy by proposing a model that integrates the Lever-arm theory with certain components of random nature. Specifically, via the comparison of our theoretical results with those yielded by the experiments, we shall test the assumption that during the time interval elapsing between ATP hydrolysis and the final release of the phosphoric radical, the position of the myosin head is determined as the result of a dynamical process that includes a macroscopic non deterministic force responsible for a non-zero average net displacement. As a consequence, the coupling between the ATP cycle and the mechanical effects should appear to be less rigid, and the actomyosin dynamics should definitely exhibit variability features of the kind mentioned in the above~\ref{Oss1}.$\div$\ref{Oss3}. items.
\par
The conjectured random dynamics above outlined will be described in details in Section 2 with a specific reference to an earlier paper~\cite{buo03}. Here we limit ourselves to showing that, within such a framework, it is possible to handle a key point of the mentioned controversy. Indeed, in~\cite{cyr00} the essential relevance of the role played by the length of the myosin head is stressed, since the Lever-arm theory makes the sliding distance less than, and somewhat proportional to, such length. This would for instance imply that reducing the length of the myosin neck to a half should reduce to a half the sliding distance, which appears to be confirmed by the experiments in~\cite{war00}. On the contrary, experiments performed by the Yanagida group only show slight changes of the overall displacement of the myosin head even after complete removal of its neck~\cite{cyr00}.
\par
In order to reconcile these evident discrepancies, we shall assume that the displacement $X$ of the myosin head during the ATP cycle can be represented 
as follows:
\begin{equation}\label{Spostamento}
X=rX_R + dX_D
\end{equation}
where $X_R$ denotes the displacement induced by the random force and $X_D$ the displacement generated by the power stroke, and where $r$ and $d$ are constants, each equal to 0 or 1. Note that the case $r=0$ and $d=1$ yields the Lever-arm theory; instead, setting $r=1$ and $d=0$ depicts a purely random situation, in the absence of any sliding due to the power stroke. Finally, the case $r=d=1$ leads to an integration of the two theories. With the choice $r=d=1$, the controversial results related to the length of the neck of myosin head can be overcome by assuming that the random displacement is a few times larger than the deterministic one. This assumption implies that only slight changes of the total distance traveled by myosin head would be observable when the contribution $X_D$ due the power stroke is made somewhat smaller by shortening the length of the myosin neck.
\section{The model}\label{model}
Let $L$ denote the distance between each pair of neighboring actin monomers. As suggested in recent literature (see~\cite{ish00} and ~\cite{kit01}) in our computations we shall take \numdim{L=5.5}{nm}. In addition, we shall assume that the magnitude of the sliding induced by the power stroke equals that of a step during the random phase, and thus take $X_D\simeq L$.
\par
Our conjectured relation~(\ref{Spostamento}) with $r=d=1$ is preliminarily supported by the bar chart in Fig. 4c) of~\cite{kit99} showing that at least one step is performed by each and every myosin head during the rising phase\footnote{The rising phase is the time interval starting with the hydrolysis of the ATP molecule and ending with the final release of the phosphoric radical Pi.}. The second column of Table~\ref{DistribuzioneNumeroNettoSalti} shows the heights of the columns of the mentioned bar chart, whereas the first column indicates net step numbers, i.e. the integral part of the ratios $X/L$. The quoted experiment was performed under low load conditions by means of microneedles having stiffness less than \numdim{0.1}{pN$/$nm}.
\par
\begin{tabMacPc}
  {\begin{tabular}{|c|r||c|r|}\hline
$\lfloor X/L \rfloor$ & Observed frequency & $\lfloor X_R/L \rfloor$ & Theoretical frequency\\ \hline\hline
1& 14 & 0 & 15 \\ \hline
2& 21 & 1 & 22 \\ \hline
3& 18 & 2 & 17\\ \hline
4& 10 & 3 &  8\\ \hline
5&  3 & 4 &  3\\ \hline
6&  0 & 5 &  1\\ \hline \hline
total&66 & total&66\\ \hline
   \end{tabular}}
   {12 cm}
   {Observed distribution of net number of steps performed by myosin heads as indicated in~\cite{kit99}. Here $\lfloor X/L \rfloor$ is the total step number during the entire rising phase, whereas $\lfloor X_R/L \rfloor\equiv\lfloor X/L \rfloor-1$ denotes the total ( algebraic) step number during the random part of the rising phase.}
   {DistribuzioneNumeroNettoSalti}
\end{tabMacPc}
As shown by the first two columns of Table~\ref{DistribuzioneNumeroNettoSalti}, one sees for instance that out of 66 observed myosin heads (all attached to a glass microneedle of stiffness less than \numdim{0.1}{pN$/$nm}) 14 performed a unit net steps, 21 a net step number equal to 2, etc., to conclude that 3 of them have performed a net step number equal to 5 implying total displacement of about \numdim{30}{nm}. Columns 3 and 4 of Table~\ref{DistribuzioneNumeroNettoSalti} show that the net step number $\lfloor X_R/L \rfloor$ of the random component is well fitted by a Poisson distribution with parameter $\hat{\eta}=1.5$ given by the ratio of the total number of performed net steps ($99$) to the number of considered myosin heads ($66$).
\par
The agreement between experimentally observed frequencies and those predicted via the Poisson distribution, jointly with the rarity of the backward steps, leads one to conclude that the \lq\lq dwell time\rq\rq\ (namely the time interval elapsing between two successive steps) should be, to a good approximation, exponentially distributed. Such conclusion is experimentally supported by the data summarized in Fig.~4b) of~\cite{kit99} leading to the estimate of about \numdim{5}{ms} for the mean dwell time. It must be pointed out that in the mentioned estimation the dwell time of all performed steps have been included, whereas our conjecture on the exponential distribution of dwell time only refers to the steps performed during the random phase. However, such a diversity does not lead to inconsistencies if one assumes that the dwell time between the last step due to the random phase and the step due to the deterministic power stroke is exponentially distributed as well. 
\par
Our facing sequences of rare events with exponentially distributed interarrival times is strongly suggestive of a first-exit problem out of an interval for continuous Markov processes possessing an equilibrium point sufficiently far from at least one of the end points of the diffusion interval (See, for instance,~\cite{nob85}).
\par
On the ground of all foregoing considerations, with reference to the random rising phase we are lead to construct a model for the actomyosin dynamics that is based on the following assumptions:
\begin{itemize}
   \item [(i)] The complex M.ADP.Pi + Energy is viewed as a point--size particle moving along an axis $X$ on which the abscissa $x$ denotes the displacement of the particle from the starting position. The positive direction of $X$ is that of the forward steps of myosin head.
   \item [(ii)] The particle is embedded in a fluid. Hence, it is not only subject to a dissipative viscous force characterized by a drag coefficient $\beta$, but also to microscopic forces originating from the thermal motion of the fluid molecules. On account of the fluctuation-dissipation theorem, such microscopic forces can be macroscopically described by means of a Gaussian white noise having intensity $2\beta$\eneter, where \cosbol\, is Boltzmann constant and $T$ the absolute temperature.
   \item [(iii)] The global interaction of the particle with the actin filament is synthesized in a conservative unique force deriving from a potential $U(x)$. The structure of the actin filament suggests that $U(x)$ be a periodic function with period equal to the distance $L$ between pairs of consecutive actin monomers: 
\begin{equation}\label{Potenziale}
U(x)=U(x+rL), \qquad \forall r\in \mathbb{Z}.
\end{equation}
Henceforth we shall denote by $L_A$ ($0<L_A<L$) the minimum of $U(x)$, assume $U(L_A)=0$ and denote by $U_0:=U(0)=U(L)$ the depth of the potential well, namely the maximum of $U(x)$.
   \item [(iv)] The particle's dynamics is described by Newton's equation in which the total acting force is the sum of two terms: the first term is a deterministic force generated by the potential $U(x)$ and by the viscous force, while the second term is the random force due to the presence of the Gaussian white noise.
    \item [(v)] One more (constant) force $F_i>0$ acts on the particle. Here we assume that this force is generated by a process that finds its origin in a part of the energy possessed by the particle, and take $F_iL\ll U_0$.
\end{itemize}
\par
Summing up, we are assuming that the complex M.ADP.Pi + Energy can be looked at as a Brownian particle subject to a tilted potential $V(x)$:
\begin{equation}\label{Potenzialeinclinato}
V(x)=U(x)-Fx \\
\end{equation}
where $U(x)$ has been indicated in (\ref{Potenziale}), $F:=F_i-F_e$, with $F_e>0$, is a constant external force, possibly applied from the outside by the experimenter. In the experimental conditions the height of the potential wells is \numdim{U_0\le 100}{pN$\cdot$nm}, the period of the potential is \numdim{L=5.5}{nm}, the particle mass is \numdim{m=2.2\cdot10^{-22}}{kg}, the drag coefficient is \numdim{\beta=90}{pN$\cdot$ns/nm} and the environmental temperature is \numdim{T=293}{K}. Therefore, Reynolds's number is much less than 1 (see, also, ~\cite{shi03}), so that the inertial term of the equation of motion can be disregarded. In conclusion, the overdamped equation describing the movement of the particle is the following Langevin equation:
\begin{equation}\label{Eq.Langevin}
\dot{x}=-\frac{1}{\beta}\,{V^\prime(x)}+\sqrt{\frac{2\mbox{\cosbol} T}{\beta}}\,\,\Lambda (t)
\end{equation}
where $\Lambda(t)$ is a zero-mean white Gaussian noise with unit intensity, ( $\dot{ }$ ) denotes time derivative and ( $^\prime$ ) space derivative.
\par
Within this framework, this idealized Brownian particle randomly moves around an equilibrium point located at a minimum $L_A$ of $U(x)$. Whenever it exits the current potential well, we conventionally say that the corresponding myosin head has made a step, in the forward or in the backward direction according to where the exit has taken place.
Hence, we are facing a first-exit problem of the Brownian particle from the endpoint of the current potential well. Taking into account the above assumptions, we conclude that the distribution of the first-exit time of the Brownian particle from the potential well is exponential, since:
\begin{itemize}
    \item [--] the process described by equations~(\ref{Potenziale}), (\ref{Potenzialeinclinato}) and~(\ref{Eq.Langevin}) possesses an equilibrium point at the minimum $L_A$ of $U(x)$;
    \item [--] the time for the particle to travel the distance $L_A$ in the presence of the only force due to $U(x)$ is (see for instance,~\cite{luc99}) $\tau=\beta L_A^2/U_0$;
    \item [--] the standard deviation of the Gaussian steady-state distribution of the process modeling the particle's motion is $\sigma=\sqrt{(\mbox{\cosbol}T/\beta)\cdot\tau/2}\equiv L_A/\sqrt{2u_0}$, where we have set $u_0=U_0 /\mbox{\cosbol}T$;
    \item [--] the ratio $l_A:=L_A/\sigma=\sqrt{2u_0}$ falls well within the interval $(2,4\sqrt{2})$ for all choices of $u_0\in[2,16]$ to which we shall refer in the foregoing, so that the exponential approximation for the first-exit time is valid~\cite{nob85}.
\end{itemize}
\par
In the sequel, we shall exploit the known formulas~\cite{lin01} for the probability $p$ that the particle exits from the current potential well to reach next well,
\begin{eqnarray}
p&=&\frac{1}{1+\exp{\left(-FL/\mbox{\cosbol}T\right)}}\label{P}.
\end{eqnarray}
and for the mean first-exit time $\mu$ from a potential well:
\begin{eqnarray}
\mu&=&\beta \frac{p}{\mbox{\cosbol}T}\int_0^Ldx\,\exp\left\{\frac{V(x)}{\mbox{\cosbol}T}\right\}\int_{x-L}^xdy\,\exp\left\{-\frac{V(y)}{\mbox{\cosbol}T}\right\}\label{Mu}.
\end{eqnarray}
\section{Determination of $F_i$ and $U_0$}
Hereafter, we shall view as constants the parameters $\beta$ and $T$ that characterize the thermal bath and the period $L$ of the potential $U(x)$, their values being specified as in Section~\ref{model}. Hence, the quantitative specification of the model described by Eqs.~(\ref{Potenziale}), (\ref{Potenzialeinclinato}) and~(\ref{Eq.Langevin}) requires that numerical values be attributed to three more parameters: the depth $U_0$ of the potential well, the position $L_A$ of the minimum of $U(x)$ in $(0,L)$ and the internal force $F_i$. The numerical specification of these parameters can be performed after the function $U(x)$ has been chosen. We shall preliminarily take $U(x)$ as a symmetric $(L_A=L/2)$ saw-tooth potential, as defined in Table~\ref{Potenziali}. Successively, we shall test the robustness of our model by assuming alternative potential functions, henceforth called \lq\lq potential profiles\rq\rq. It should be noted that $F_i$ is somewhat related to the largest force that myosin is able to endogenously generate. Here, we shall not attempt to provide any biological justification of its genesis. We limit ourselves to pointing out that elsewhere~\cite{nis02}, where similar experiments were performed on Myosin~VI, it is conjectured that the week binding between actin and myosin is a source of distortion of the geometry of the two helixes in the actin filament. Such a distortion exposes the hydrophobic region of actin to myosin head, thus generating a tilt of the potential, and hence a constant force $F_i$. The existence of a tilt of the potential, conjectured in~\cite{buo03}, has been successively supported by means of simulations in~\cite{esa03} where it is shown that in the absence of such a tilt experimental available evidence on the myosin motion in the presence of contrasting applied loads cannot be accounted for by any of the other models therein considered. Within our strictly phenomenological framework the existence of this force $F_i$ is supported by Eq.~(\ref{P}) showing that in the absence of external applied forces (i.e. $F_e=0$) steps on either directions would be equally likely unless $F_i\ne0$, in contrast with the experimental evidence on the high degree of directionality exhibited by motion of the myosin head.
\par
To proceed along the quantitative specification of our model in a way to be able to attempt the fitting of available experimental data, the values of $F_i$ and of $U_0$ must be specified. This will be done by making use of the available experimental data~\cite{kitpr} shown in Table~\ref{Steps} and in Table~\ref{DwellTime}\footnote{Note that loads and dwell times in Table~\ref{DwellTime} must be viewed as averages to which confidence intervals are associated. For instance, the load \numdim{0.046}{pN} is the result of all measurements for which the product of the stiffness of the glass microneedle times the distance traveled by the myosin head falls around \numdim{0.046}{pN}. The corresponding dwell time \numdim{5.3}{ms} is to be viewed as the arithmetic average of the dwell times recorded during these measurements.}.
\begin{table}[htb]
        \begin{minipage}[htb]{9 cm}
\renewcommand{\arraystretch}{2.5}
            \caption{For three conditions of the applied load $C$ the recorded numbers $\hat{n}_f$ of forward steps, $\hat{n}_b$ of backward steps, the total number of steps and the percentage $\hat{p}$ of forward steps are listed.}
            \label{Steps}
            \vspace{0.2 cm}
            \begin{center}
            \begin{tabular}{|c|c|c|c|c|}\hline
            $C$ (pN) & $\hat{n}_f$ & $\hat{n}_b$ & $\hat{n}_f+\hat{n}_b$ & $\displaystyle{\hat{p} = \frac{\hat{n}_f}{\hat{n}_f+\hat{n}_b}}$\\ \hline
            $\big]0.0,0.5\big]$ & 54  &  9 & 63 & 0.8571 \\\hline
            $\big]0.5,1.0\big]$ & 40  &  9 & 49 & 0.8163 \\ \hline
            $\big]1.0,2.0\big]$ & 29  & 19 & 48 & 0.6042 \\ \hline
            \end{tabular}
            \end{center}
        \end{minipage}
    \hfill
        \begin{minipage}[htb]{5 cm}
            \caption {Recorded dwell times $\hat{\mu}$ for different values of the applied load $C$.}
            \label{DwellTime}
            \vspace{0.2 cm}
            \begin{center}
            \begin{tabular}{|r|r|}\hline
            $C$ (pN) & $\hat{\mu}$ (ms) \\ \hline
            0.046 &  5.3 \\ \hline
            0.190 &  5.7 \\ \hline
            0.300 &  6.0 \\ \hline
            0.470 &  7.1 \\ \hline
            0.690 &  8.9 \\ \hline
            0.830 &  6.2 \\ \hline
            1.240 & 11.1 \\ \hline
            1.890 & 11.0 \\ \hline
            \end{tabular}
            \end{center}
    \end{minipage}
\end{table}
From them it is evident that the frequency $\hat{p}$ of forward steps decreases as the applied load increases, while the dwell times increase with the load. Since the external space-dependent force accounting for the applied load is not included in our above formulated model, one more assumption is necessary: 
\begin{itemize}
   \item [(vi)] The external constant force $F_e$ acting on the myosin head equals the value $C$ obtained by the elastic force $C(x)$ (given by the product of the microneedle stiffness times the distance traveled by the myosin head) at the end of the rising phase and its direction is opposite to that in which motion occurs.
\end{itemize}
In other words, in our model the presence of applied loads in the experimental setup is expressed by means of the variable $F_e$.
\par
For some fixed values of internal force $F_i$, use of Equation~(\ref{P}) has been made to calculate the theoretical probabilities of the particle's exit from the current potential well to the next well (i.e. the analogue of the forward step probabilities) as function of $F_e$. The results are shown in Figure~\ref{PversusFE}, where eight realistic values of $F_i$ have been chosen in the interval \numdim{\big[1.00}{pN}\numdim{,1.90}{pN}$\big]$. Vertical lines indicate the $3$ load intervals of Table~\ref{Steps}, whereas horizontal lines indicate the $3$ corresponding recorded frequencies. We see that for internal forces \numdim{F_i=1.00}{pN} and \numdim{F_i=1.90}{pN} the plotted curves do not meet the requirement of leading to the experimentally recorded frequency $\hat{p}=0.8571$, whatever value $F_e$ is chosen within interval \numdim{\big[0}{pN}\numdim{,0.50}{pN}$\big]$. Similarly, for \numdim{F_i=1.55}{pN} no value of the computed probability equals the frequency $\hat{p}=0.8163$ for $F_e$ ranging in \numdim{\big[0.50}{pN}\numdim{,1.00}{pN}$\big]$. Instead, all remaining $5$ curves referring to values of $F_i$ ranging from \numdim{1.60}{pN} to \numdim{1.80}{pN} in steps of magnitude \numdim{0.05}{pN}, from below upward, are in agreement with the experimental values of Table~\ref{Steps}. Hence, the interval of values for $F_i$ to be selected in order to secure the fitting of the experimental data is \numdim{\big[1.60}{pN}\numdim{,1.80}{pN}$\big]$.
\par
\begin{figMacPc}
   {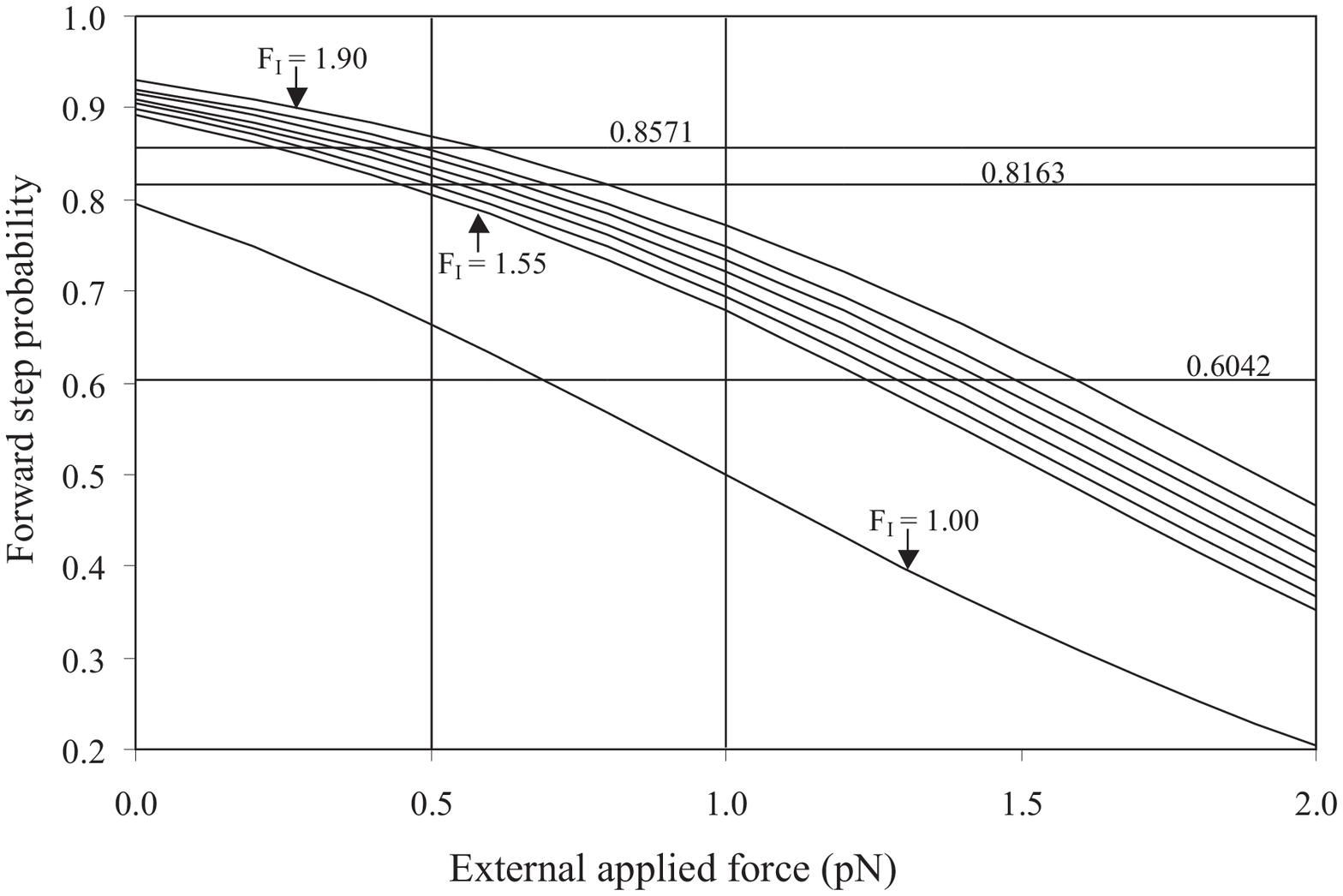}
   {13cm}
   {The probability $p$ that the particle exits from the current potential well to reach next potential well is plotted as function of the external applied force $F_e$ for various values of internal force $F_i$~(pN). Temperature and potential period are taken as \numdim{T=293}{K} and \numdim{L=5.5}{nm}, respectively.}
    {PversusFE}
\end{figMacPc}
We now come to the estimation of the last parameter, $U_0$, i.e. of the depth of the potential well. This is done by exploiting Eq.~\ref{Mu} in which the left-hand side is viewed as the function $\mu=\mu(F_i,F_e,U_0;L,\beta,T)$ and fixed to \numdim{5.3}{ms}, that (see Table~\ref{DwellTime}) corresponds to the smallest applied load \numdim{0.046}{pN}\footnote{We choose the smallest load to minimize the error due to the approximation of elastic force $C(x)$ by $F_e$.}. Since temperature $T$, period $L$ of the potential $V(x)\equiv U(x)-(F_i-F_e)x$, and drag coefficient $\beta$ are specified, if we take \numdim{F_e=0.046}{pN} Eq.~(\ref{Mu}) makes $U_0$ an implicit function of $F_i$.
\par
Note that as $F_i$ increases (i.e. as the potential tilt increases) the mean first-exit time $\mu$ decreases. Hence, in order to keep $\mu$ constantly equal to \numdim{5.3}{ns}, while $F_i$ ranges in the interval \numdim{\big[1.60}{pN}\numdim{,1.80}{pN}$\big]$, depth $U_0$ must be taken as a monotonically increasing function of $F_i$. From $\mu(1.60,0.046,U_0)=5.3$ and $\mu(1.80,0.046,U_0)=5.3$, Eq.~(\ref{Mu}) yields \numdim{U_0\approx15.632}{\eneter} and \numdim{U_0\approx 15.755}{\eneter}, respectively.
\par
In order to test the agreement of our model with the experimental values of mean dwell times for the various loads (see Table~\ref{DwellTime}), we make use of Eq.~(\ref{Mu}) to determine $\mu$ as a function of $F_e$ for the pairs (\numdim{1.60}{pN},\numdim{15.632}{\eneter}) and (\numdim{1.80}{pN},\numdim{15.755}{\eneter}), involving the extrema of the determined values for $F_i$ and $U_0$. The obtained values are showing Figure~\ref{MuversusFE}, where the corresponding experimental mean dwell times are also indicated. Obviously, all other pairs of admissible values of $F_i$ and $U_0$ lead to curves lying inside of the above two pairs. 
\begin{figMacPc}
   {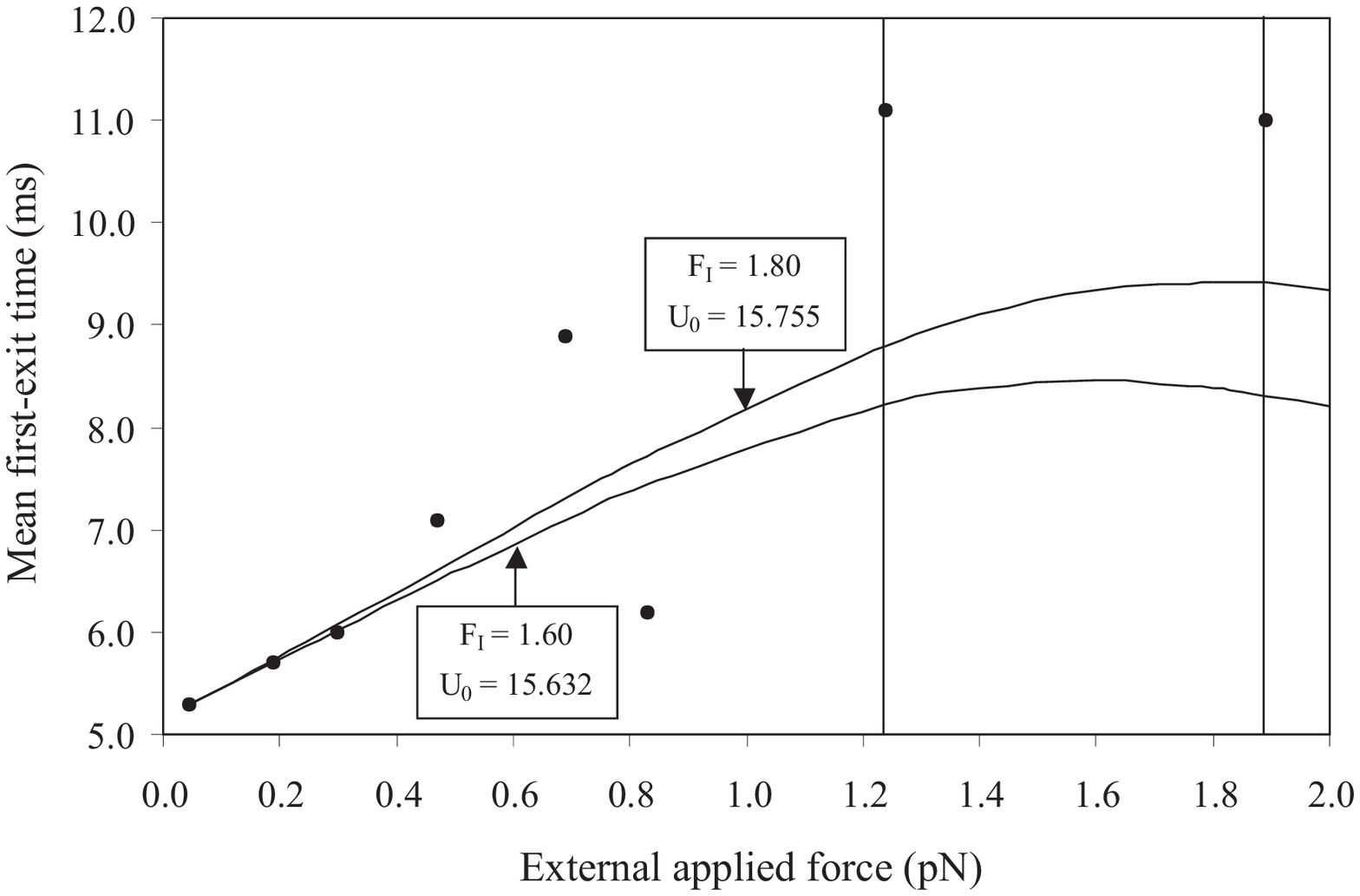}
   {13cm}
   {First-exit time $\mu$ is plotted as a function of the external applied force $F_e$. The chosen values of $F_i$~(pN) and $U_0$~(\eneter) are indicated for each curve. Dots represent the experimental dwell times of Table~\ref{Steps}. Drag coefficient $\beta$ has been taken as \numdim{90}{pN$\,$ns$/$nm}, whereas temperature and period of the potential are the same as in Figure~\ref{PversusFE}.}
   {MuversusFE}
\end{figMacPc}
Inspection of Figure~\ref{MuversusFE}, jointly with the magnitudes of the confidence intervals associated to each experimental value~\cite{kit01}, shows that for small values of $F_e$ the agreement between experimental data and theoretical predictions is good. Indeed, up to the first $3$ values of Table~\ref{Steps} the agreement is excellent, to become more than acceptable for larger values of $F_e$ up to \numdim{0.83}{pN}. The discrepancies shown by the remaining two experimental data are a  consequence of the crude assumption (vi) above by which the applied load is assumed to be strictly parallel to the direction of motion. This is acceptable for reasonably small loads, but certainly unrealistic for large loads: In this case a significant orthogonal component should be expected, whose effect is alike to an increase of the depth of the potential well $U_0$. Nevertheless, in the interval between the last two load values of Table~\ref{Steps}, a qualitative similar behavior of theoretical curves and experimental values is present.
\section{Net step number distribution}
In the present Section we shall implement the model described by Eqs.(\ref{Potenziale})--(\ref{Eq.Langevin}) in order to obtain, via a suitable simulation procedure, the distribution of the net number of steps performed by the myosin head during the time interval elapsing between the instant when the ATP molecule is hydrolyzed and the final release of the phosphoric radical, i.e. during the random part of the rising phase. The results of our simulations will then be discussed with reference to the experimental data of Table~\ref{DistribuzioneNumeroNettoSalti}. We recall that such data refer to 66 myosin heads that have globally performed 99 net steps.
\par
Our simulation procedure is based on a discretized version of Eq.~(\ref{Eq.Langevin}) and on a routine for generating Gaussian pseudorandom numbers in a way to determine step by step the positions achieved by 66 Brownian particles all originating at $x_0$ at time 0. The traveled distances at the end of the individual random rising phases are recorded, which finally leads one to the determination of the 66 net step numbers. Such procedure has been implemented on an IBM SP4 parallel supercomputer and repeated 1600 times to obtain a reliable statistics.
\par
In order to solve Eq.~(\ref{Eq.Langevin}) numerically, one has to specify preliminarily initial position $x_0$ of the Brownian particle and duration $\Theta$ of each sample path. We shall safely take $x_0=L/2$ since, whatever the actual initial position of the particle inside of the potential well, the its relaxation time is much smaller than the mean first-exit time. The specification of sample path duration is somewhat more involved because $\Theta$ is a random variable of unknown distribution. Nevertheless, we can appeal to the approximate exponential distribution of the first-exit time, motivated by the depth of the potential well, to assume that $\Theta$ is approximately gamma-distributed, with probability density $\Gamma(\mu,\nu)$ where $\mu$ is the mean first-exit time from a potential well and $\nu$ is the mean total exits of the Brownian particle. While $\mu$ is obtained via Eq.~(\ref{Mu}), our estimate $\hat{\nu}$ of $\nu$ is obtained by using the data of Table~\ref{DistribuzioneNumeroNettoSalti} and Eq.~(\ref{P}):
\begin{displaymath}
\hat{\nu} = \frac{99}{66(2p-1)}.
\end{displaymath}
\par
Finally, for each specified potential well $U_0$, the size $\Delta t$ of the time parsing in Eq.~(\ref{Mu}) is determined by progressively reducing it until the obtained distribution becomes appreciably invariant. For an immediate comparison, Table~\ref{pot.cinque} shows the distribution obtained via simulation for a potential \numdim{U_0=5}{\eneter} and a parsing time \numdim{\Delta t=0.25}{ns}\ jointly with the experimental distribution. The agreement between observed and numerical frequencies is more than satisfactory. 
\par 
\begin{tabMacPc}
   {\begin{tabular}{|c|r||r||r||r|}\hline
$\lfloor X/L \rfloor$ &Observed frequency & \multicolumn{3}{c|}{Numerical frequency} \\ \hline
& &\numdim{F_i=1.70}{pN} &\numdim{F_i=1.75}{pN} &\numdim{F_i=1.80}{pN}\\ \hline\hline
-1& 0 & 2.05 & 1.90 & 1.76 \\ \hline
0& 14 &13.79 &13.63 &13.88 \\ \hline
1& 21 &20.72 &20.83 &20.93 \\ \hline
2& 18 &16.41 &16.43 &16.47 \\ \hline
3& 10 & 8.53 & 8.59 & 8.54 \\ \hline
4&  3 & 3.14 & 3.26 & 3.16 \\ \hline
5&  0 & 0.93 & 0.93 & 0.88 \\ \hline \hline
\multicolumn{2}{|r||}{$\overline{CV}$}       &    0.966  &    0.964  &    0.963 \\ \hline
\multicolumn{2}{|r||}{$\overline{p}$}        &    0.909 &    0.914 &    0.920\\ \hline
\multicolumn{2}{|r||}{$p$}          &    0.910 &    0.915 &       0.920\\ \hline
\multicolumn{2}{|r||}{$\overline{\mu}$ (ns)} & 1235.057 & 1207.288 & 1180.852\\ \hline
\multicolumn{2}{|r||}{$\mu$ (ns)}            & 1233.784 & 1206.048 & 1178.695\\ \hline
   \end{tabular}}
   {14 cm}
   {Net step frequency distributions for \numdim{U_0=5}{\eneter} and for three values of $F_i$ chosen in the admissible interval. Here $F_e=0$, $x_0=L/2$ and \numdim{\Delta t=0.05}{ns}. Other parameters have been chosen as in Figure~\ref{MuversusFE}. The last 5 rows show the coefficient of variation, probability of a step forward and mean exit time, all obtained via simulations. The theoretical values of $p$ and $\mu$, obtained via Eqs.~(\ref{P}) and (\ref{Mu}), are also indicated.}
   {pot.cinque}
\end{tabMacPc}
The chosen value of $U_0$ is motivated by a twofold consideration. First of all, it is large enough to imply that the first-exit time distribution is approximately exponential. This is indeed supported by the numerically evaluated variation coefficient that has been found to be 0.96. Second, and most relevant consideration, is that it is reasonable to conceive that, under the assumed overdamped regime, the net step number is insensitive to the depth of the potential well. Indeed, the forward exit probability $p$ given by~(\ref{P}), under the fixed environmental temperature, only depends on the energy $FL$, namely on the difference of potential $V(x)$ over one period, thus being independent of the depth $U_0$. Table~\ref{pot.quattro&otto} evidently supports such conclusion. Indeed, it indicates that, for instance, by doubling the depth of the potential well $U_0$, the net step frequencies are not affected significantly, even for different choices of internal forces\footnote{The presence of non-integer numbers in Table~\ref{pot.cinque}, ~\ref{pot.quattro&otto} and ~\ref{pot.cinquebis} is a consequence of our adopted estimation procedure. The half-width of the related 95\%-confidence intervals has been seen never to exceed $0.2$. For comparison reasons we have not rounded out the raw numbers to the nearest integers.}.
\begin{tabMacPc}
   {\begin{tabular}{|c||r|r||r|r||r|r|}\hline 
    & \multicolumn{2}{c||}{\numdim{F_i=1.70}{pN}}&\multicolumn{2}{c||}{\numdim{F_i=1.75}{pN}}    & \multicolumn{2}{c|} {\numdim{F_i=1.80}{pN}} \\ \hline
  $\lfloor X_A/L \rfloor$ & \numdim{U_0=4}{\eneter} & \numdim{U_0=8}{\eneter} & \numdim{U_0=4}{\eneter} & \numdim{U_0=8}{\eneter} & \numdim{U_0=4}{\eneter} & \numdim{U_0=8}{\eneter} \\ \hline
\hline -1&  2.11 &  2.05 &  1.94 &  1.90 &  1.85 &  1.81 \\
\hline  0& 13.56 & 13.96 & 13.54 & 13.76 & 13.48 & 14.00 \\
\hline  1& 21.13 & 20.26 & 20.92 & 20.26 & 21.30 & 20.46 \\
\hline  2& 16.70 & 16.11 & 16.93 & 16.29 & 16.98 & 16.09 \\
\hline  3&  8.43 &  8.58 &  8.53 &  8.61 &  8.40 &  8.56 \\
\hline  4&  2.95 &  3.40 &  3.01 &  3.53 &  2.91 &  3.49 \\
\hline  5&  0.75 &  1.12 &  0.78 &  1.15 &  0.76 &  1.11 \\
\hline
   \end{tabular}}
   {16 cm}
   {Net step frequency distribution. Here $F_e=0$ and $x_0=L/2$. Other parameters are chosen as in Figure~\ref{MuversusFE}. Parsing steps are chosen as follows: \numdim{\Delta t=0.1}{ns} for \numdim{U_0=4}{\eneter}, and \numdim{\Delta t=0.025}{ns} for \numdim{U_0=8}{\eneter}. The indicated values of $F_i$ belong to the admissible interval.}
   {pot.quattro&otto}
\end{tabMacPc}
\par
It should be explicitly remarked that the parsing parameter $\Delta t$ must be determined with specific reference to the magnitude of $U_0$. Indeed $\Delta t$ must be much smaller than the relaxation time of the force $-U^\prime(x)$ that is proportional to $\beta L^2/U_0$. Furthermore, it must be such as to cope with the random forces due to the thermal bath. Therefore, the mean square displacement per unit time should remain constant as the depth $U_0$ of $U(x)$ is made to change, which implies an inverse dependence of the magnitude of $\Delta t$ on the square of magnitude of the potential well. This is the motivation for the indicated choices of $\Delta t$ in Table~\ref{pot.quattro&otto}.
\section{The role of potential forms and asymmetries}
\begin{figure}[htb]
      \begin{minipage}[htb]{7.5 cm}
          \epsfxsize=7 cm
          \centerline{\epsfbox{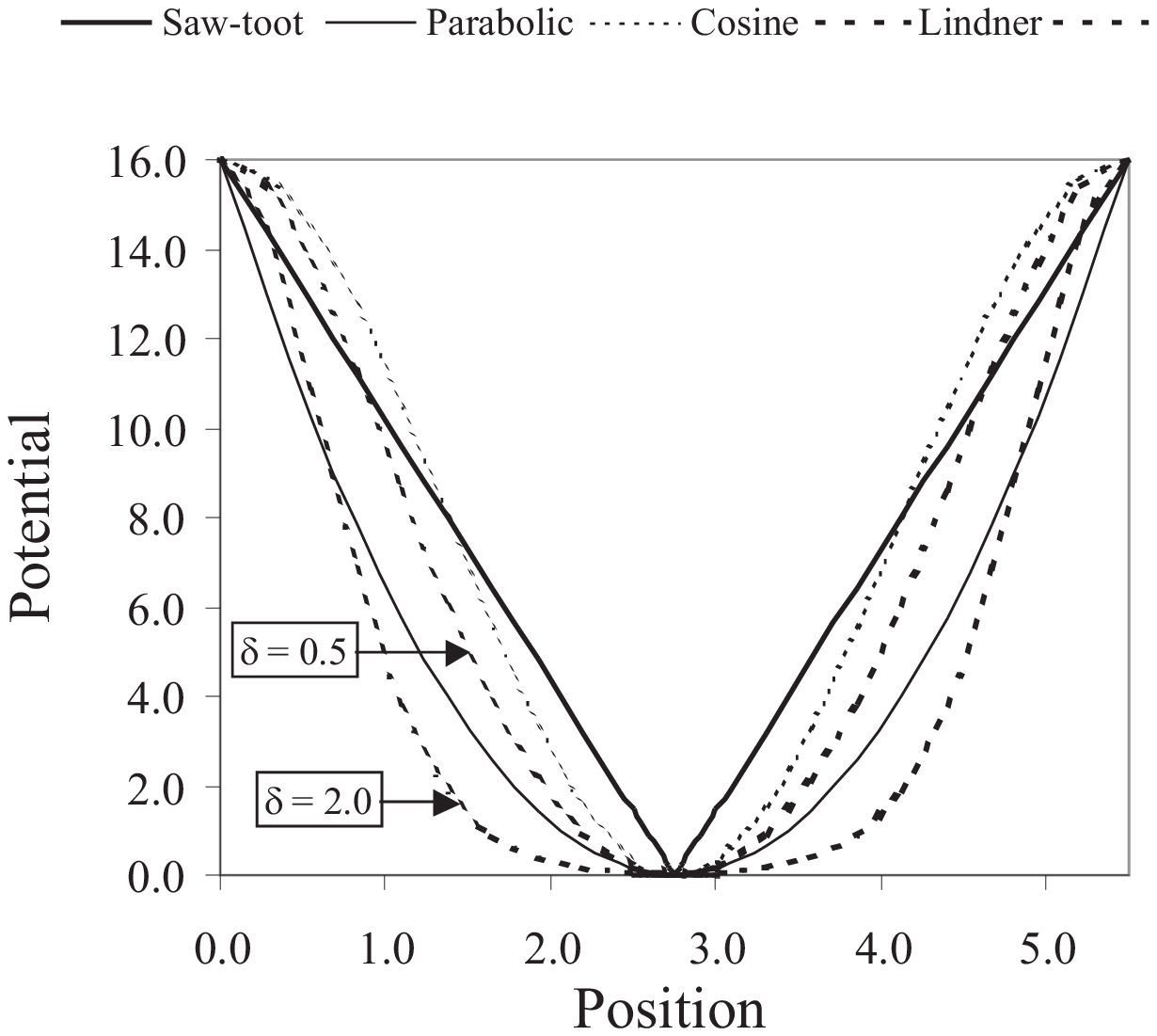}}
          \caption{Plots of the potentials as function of the position within each pair of consecutive monomers.}   
          \label{GraficiPotenziali}
      \end{minipage}
    \hfill
       \begin{minipage}[htb]{7.5 cm}
          \epsfxsize=7 cm
          \centerline{\epsfbox{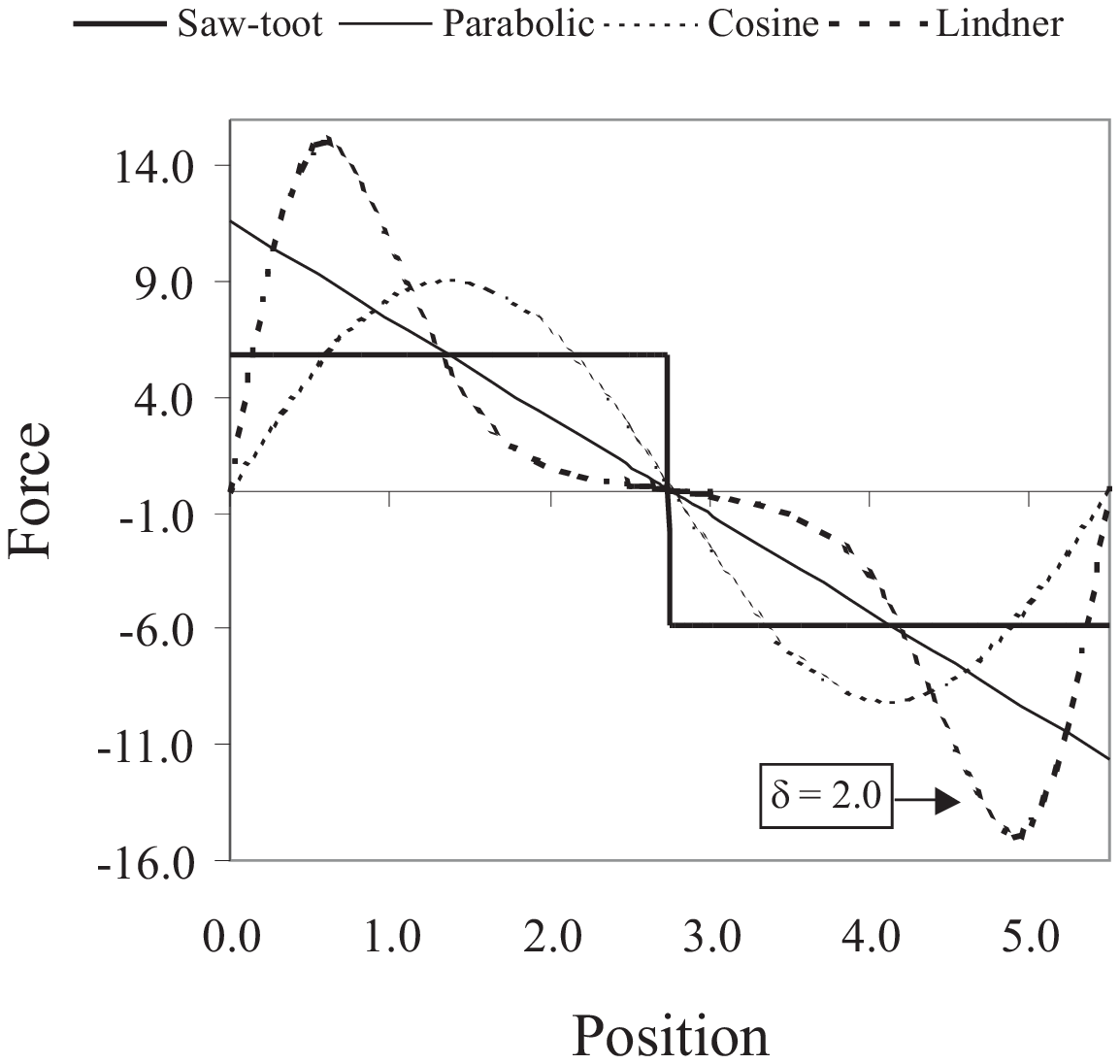}}
          \caption{Conservative forces originated by the potentials of Figure~\ref{GraficiPotenziali}.}   
          \label{GraficiForze}
      \end{minipage}
\end{figure}

In~\cite{buo03} a preliminary version of the model considered here was discussed with reference of a parabolic potential, henceforth called \lq\lq parabolic profile\rq\rq. The evaluation of the robustness of the model was, however, postponed to a successive investigation. This task is the object of the present Section in which, in addition to the case of saw-toot profile, three more profiles will be considered: parabolic, cosine-like and Lindner-type. (See Table~\ref{Potenziali}). A plot of the considered potentials over one period are shown in Figure~\ref{GraficiPotenziali}, whereas Figure~\ref{GraficiForze} refers to the corresponding generated forces. Lindner-type potential has been indicated for two values of parameter $\delta$. It is not difficult to see that $\delta\to 0$ yields the cosine profile, whereas the potential flattens down in the middle as $\delta$ increases.
\par
\begin{table}[htb]
        \begin{minipage}[htb]{9.7 cm}
            \renewcommand{\arraystretch}{2.0}
            \caption{Potentials' profiles.}
            \label{Potenziali}
            \vspace{0.2 cm}
            \begin{center}
            \begin{tabular}{|l|c|}\hline
            Saw-toot    & $U_S(x)$  $=\left\{\begin{array}{l}
\displaystyle{\frac{-U_0}{L_A}\left(x-L_A\right)},\quad 0 \le x\le L_A\\     		    
\displaystyle{\frac{U_0}{L-L_A}\left(x-L_A\right)},\quad L_A \le x \le L\\
\end{array}\right.$ \\ \hline
            Parabolic & $U_P(x)$  $\displaystyle{=\frac{U_0}{L^2/4}\left(x-\frac{L}{2}\right)^2}$ \\ \hline
            Cosine     & $U_C(x)$  $\displaystyle{=\frac{U_0}{2}\left[\cos\left(\frac{2\pi}{L} x\right)+1\right]}$ \\ \hline
            Lindner & $U_L(x)$  \makebox{$\displaystyle{=\frac{U_0}{e^{2\delta}-1}}\left\{e^{\displaystyle{\delta\left[\cos\left(\frac{2\pi}{L} x\right)+1\right]}}-1\right\}$\vspace{0.2 cm}} \\ \hline
            \end{tabular}
            \end{center}
        \end{minipage}
    \hfill
        \begin{minipage}[htb]{5 cm}
\renewcommand{\arraystretch}{1.0}
            \caption{For each potential profile the depth $U_0$ of the potential well is indicated. Here \numdim{F_i=1.75}{pN}, \numdim{F_e=0.046}{pN} and the other parameters are the same as in Figure~\ref{MuversusFE}.}
            \label{Altezze}
            \vspace{0.2 cm}
            \begin{center}
            \begin{tabular}{|l|c|}\hline
            Type                   & $U_0$ (\eneter)  \\ \hline
            Saw-toot                 & 15.723 \\ \hline
            Parabolic              & 15.043 \\ \hline
            Cosine                 & 13.944 \\ \hline
	    Lindner ($\delta=0.1$) & 13.942 \\ \hline
            Lindner ($\delta=0.5$) & 13.918 \\ \hline
            Lindner ($\delta=1$)   & 13.851 \\ \hline
            Lindner ($\delta=2$)   & 13.697 \\ \hline
            Lindner ($\delta=3$)   & 13.611 \\ \hline
            Lindner ($\delta=4$)   & 13.591 \\ \hline
            Lindner ($\delta=5$)   & 13.604 \\ \hline
            Lindner ($\delta=10$)  & 13.749 \\ \hline
            \end{tabular}
            \end{center}
        \end{minipage}
\end{table}
The independence of the exit probability $p$ of the potential's profile implies that the net step distribution is profile-independent as well. This is also evident from Table~\ref{pot.cinquebis}, where the net step distributions are reported for each of the above-considered four potential profiles. 
\begin{tabMacPc}
   {\begin{tabular}{|c|r||r||r||r||r|}\hline
$\lfloor X/L \rfloor$ &Observed frequency & \multicolumn{4}{c|}{Potentials' profiles} \\ \hline
& &Saw-toot &Parabolic  &Cosine &Lindner ($\delta=2$)\\ \hline\hline
-1 &0   & 1.89 & 1.92 & 1.90 &1.99 \\ \hline
 0 &14  &13.61 &13.69 &13.73 &13.81 \\ \hline
 1 &21  &20.67 &20.69 &20.59 &20.46 \\ \hline
 2 &18  &16.47 &16.42 &16.17 &16.20 \\ \hline
 3 &10  & 8.63 & 8.60 & 8.70 &8.64 \\ \hline
 4 & 3  & 3.33 & 3.29 & 3.38 &3.34 \\ \hline
 5 & 0  & 0.98 & 0.95 & 1.05 &1.07 \\ \hline \hline
\multicolumn{2}{|r||}{$\overline{CV}$ }  &0.964 &0.970 &0.984 &0.982 \\ \hline
\multicolumn{2}{|r||}{$\mu$ (ns) }& 1206.048 & 1403.504 & 2194.007 & 2270.677\\ \hline
   \end{tabular}}
   {14 cm}
   {The first two columns list the net number of steps performed by myosin heads and the corresponding observed frequencies. The remaining four columns show the net step distributions, obtained via numerical simulations using the indicated potential profiles all possessing \numdim{U_0=5}{\eneter}. In all cases \numdim{F_i=1.75}{pN}, $F_e=0$, $x_0=L/2$, and \numdim{\Delta t=0.05}{ns}. Other parameter have been chosen as in Figure~\ref{MuversusFE}. Variation coefficients and mean-exit times $\mu$, obtained via Eq.~(\ref{Mu}), are listed as well. }
   {pot.cinquebis}
\end{tabMacPc}
\par
Next task is to pinpoint the effects of the potential profiles on the mean first-exit time. To this purpose, we refer to Table~\ref{DwellTime} showing that the mean dwell time in lowest load condition is \numdim{\mu=5.3}{ms}. After choosing \numdim{F_i=1.75}{pN}, we then make use of Eq.(\ref{Mu}) for each and every one of the four considered potentials imposing that the left hand side equals such value of $\mu$. By iterated numerical integrations, for each potential profile the corresponding value of $U_0$ is finally determined. The result are listed in Table~\ref{Altezze}, where the effect of the parameter $\delta$ in Lindner-type profile has been detailed.
\par
The behavior of mean first-exit time $\mu$ as a function of the magnitude of the external force is shown in Figure~\ref{MuversusFE+Profili}. All considered cases of Table~\ref{Altezze} lead to graphs falling in the region bounded by the lowest and the highest curves. We point out that changing the potential profiles never yields mean first-exit time changes exceeding ten percent. Hence, we are led to conclude that the depth $U_0$ of the potential well can be tuned to acceptable biological values by a suitable selection of the potential profile, without affecting appreciably the value of the mean first-exit time. For instance, switching from saw-toot to Lindner-type potential with $\delta=2$, lowers $U_0$ by more than \numdim{2}{\eneter}. (See Table~\ref{Altezze}). It is thus conceivable that a variety of potential profiles exists such that the height of the potential well can be further lowered, in a way to switch from about \numdim{15}{\eneter} as indicated in~\cite{esa03} to about \numdim{5}{\eneter} as suggested in~\cite{luc99}.
\begin{figMacPc}
   {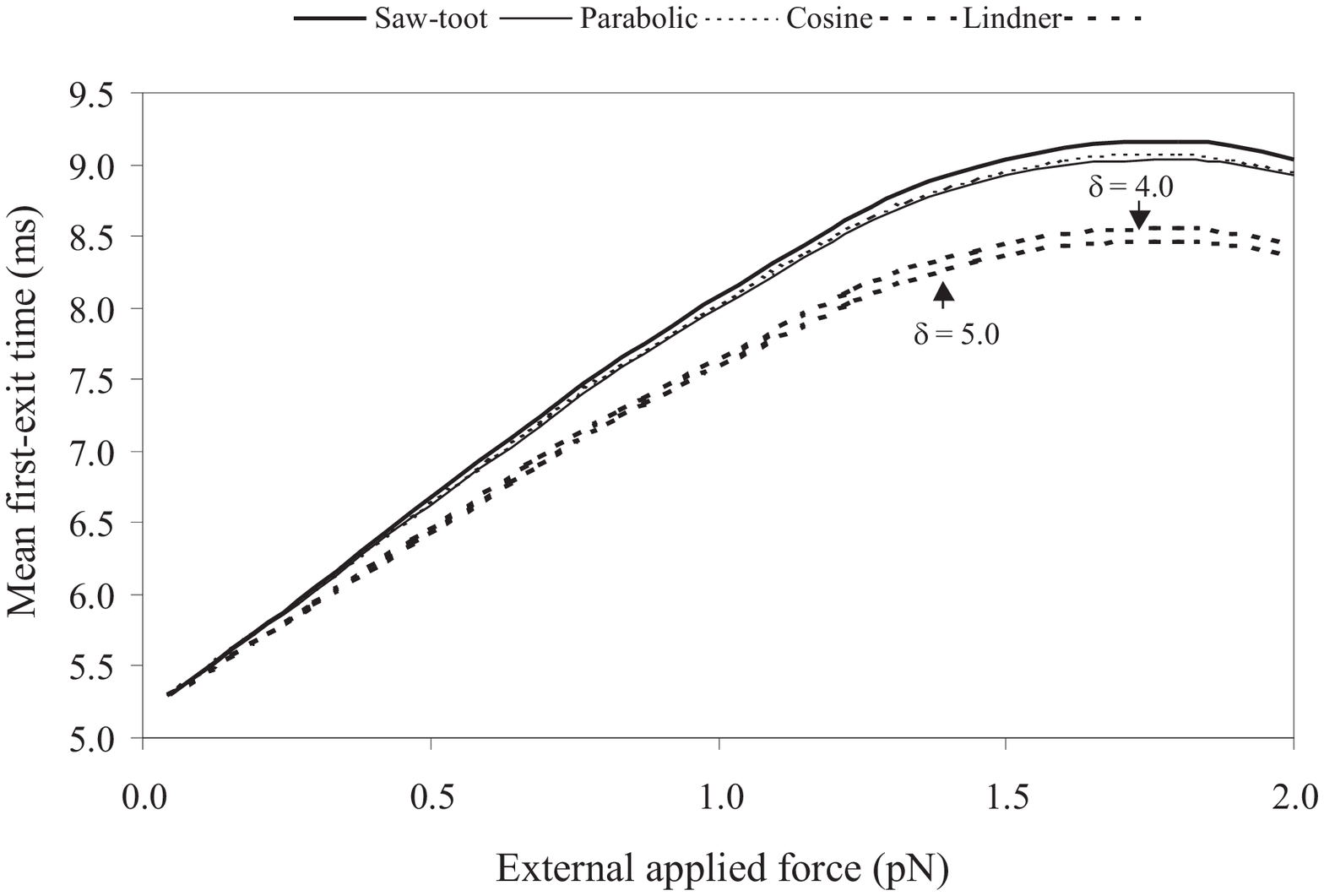}
   {13cm}
   {For each potential profile the mean first-exit time $\mu$ is plotted as a function of the external force $F_e$. The corresponding values of $U_0$, for each potential, are listed in  Table~\ref{Altezze}. We have taken \numdim{F_i=1.75}{pN}, while the other parameters have been chosen as in Figure~\ref{MuversusFE}.}
   {MuversusFE+Profili}
\end{figMacPc}
The quantitative analysis and comparison with available data has been performed under the assumption of rigorous symmetries exhibited by the profiles of the potentials generating the periodic force acting on the Brownian particles. It is, however, presumable that the complex biological reality underlying the observed motion of the myosin heads may require to relax the rigorous symmetry assumption. To test the effect of symmetry breaking, we take into consideration the saw-toot potential $U_S(x)$ of Table~\ref{Potenziali} and make it asymmetric by taking $L_A\ne L/2$, namely by shifting in either direction the point of minimum of the potential. Thus doing, the introduced asymmetry should affect the motion of the Brownian particles. While the probabilities of exit from the current potential well are insensitive to the potential's profile, and hence also to its asymmetry, the mean first-exit time $\mu$ is clearly affected by it, as shown by Eq.~(\ref{Mu}). The quantitative dependence of $\mu$ on the potential's asymmetry is indicated in Figure~\ref{MuversusFE+Asimmetria}, where on the abscissa the external force $F_e$ is indicated. Each curve is characterized by two parameters: the point $L_A$ of minimum and the depth $U_0$ of the potential. 
\begin{figMacPc}
   {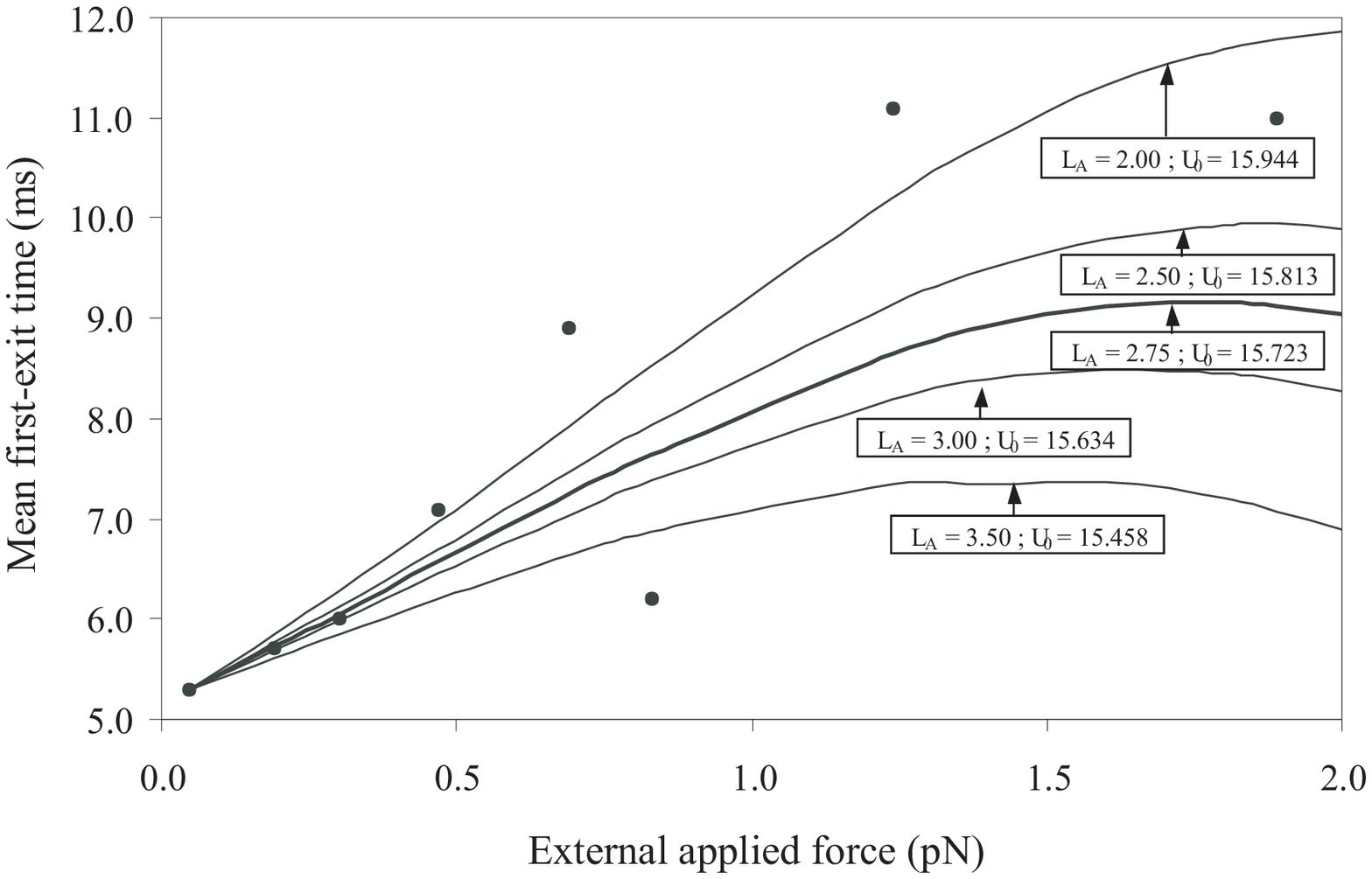}
   {13cm}
   {For the saw-toot potential profile with the indicated value of asymmetry $L_A$, the mean first-exit time $\mu$ is plotted as a function of the external force $F_e$. The corresponding values of $U_0$ such that one obtains \numdim{\mu=5.3}{ms} for \numdim{F_e=0.046}{pN}, are also indicated on each curve. We have taken \numdim{F_i=1.75}{pN}, while the other parameters have been chosen as in Figure~\ref{MuversusFE}.}
   {MuversusFE+Asimmetria}
\end{figMacPc}
From top to bottom, the first curve is characterized by asymmetry \numdim{L_A=2.0}{nm}, implying a somewhat steeper rise of the potential in the backward direction. Such asymmetry is mitigated in the next curve (\numdim{L_A=2.5}{nm}). The third curve, indicated in bold, is used as a comparison tool, as it refer to symmetric profile ($L_A=L/2$). The remaining two curves are labeled by \numdim{L_A=3.0}{nm} and \numdim{L_A=3.5}{nm}, thus representing mirror-image situation with respect to the zero-asymmetry case. Now the forward edges of the potentials are steeper. For each curve, the indicated value of $U_0$ has been determined by imposing that \numdim{\mu=5.3}{ms} when \numdim{F_e=0.046}{pN} (lowest load), so that all curves originate at the same point. As Figure~\ref{MuversusFE+Asimmetria} shows, changes of the potential's asymmetry of 30$\%$ in either direction do not affect greatly the magnitude of the mean first-exit time. Finally, we remark that suitable choices of backward asymmetry (such \numdim{L_A=2.0}{nm} in Figure~\ref{MuversusFE+Asimmetria}) improve the fitting of the experimental data even for larger load values.
\section{Concluding remarks}
The model of actomyosin dynamics discussed in the foregoing rests on the assumption that the total energy made available to the myosin head by the ATP molecule hydrolysis and by the thermal bath has a two-fold overall role: To produce the power stroke predicted by the lever-arm model and also to generate the kind of sliding of myosin head on the actin filament in the \lq\lq loose coupling\rq\rq\ mechanism originally hypothesized in~\cite{oos86} and then experimentally demonstrated in~\cite{kit99}. This is expressed mathematically via the representation of the displacement of a particle consisting of a combination of a deterministic part and of a random component. The latter is generated by the simultaneous presence of a washboard-type potential and a random force arising from thermal fluctuations. Our model has then been tested by making use of a set of data on the dwell times and step frequencies of myosin heads under various load conditions. We have shown that by a suitable tuning of the internal force and depth of the potential well, the theoretically calculated probability $p$ and mean first-exit time $\mu$ of the representative Brownian particle are in good agreement with their biological counterpart. A second set of experimental data concerning the net step number distribution of myosin heads under low load conditions has then been exploited to show that the washboard potential used by us is able to reproduce such distribution within the mathematical analogy. To the rather small size of the experimental sample should be ascribed the 3$\%$ discrepancy represented by the two net backward displacements predicted by our model.
\par
Next, the robustness of our model has been tested by inserting 4 different potential profiles in the Langevin equation of motion. The consequently performed calculations have shown that the mean exit-time depends on the external force in a way that is essentially insensitive to the chosen potential's profile. The chosen profile, instead, has been seen to play an essential role in that it critically relates the depth of the potential well to magnitude of the mean exit-time. A finer tuning of the mean exit-time is finally achieved by regulating the level of asymmetry of the potential's profile.
\acknowledgment{We thank Consorzio Universitario CINECA for providing the computation time on IBM-SP4 supercomputer.}
\vspace{10cm}

%
\end{document}